\documentclass[showpacs,preprintnumbers,amsmath,amssymb]{revtex4}

\begin{document}


\title{Comment to the paper "Radiation induced by relativistic electrons propagating through
random layered stacks: Numerical simulation results" by
A.A.Varfolomeev and et al NIM B {\bf 256}, 705 (2007)}

\author{ Zh.S. Gevorkian$^{1,2,3,*}$, J.Verhoeven$^{4}$}
\affiliation{ $^{1}$ Institute of Physics, Academia of Sinica,
Nankang, Taipei 11529, Taiwan \\
$^{2}$ Institute of Radiophysics and Electronics,Ashtarak-2,378410,Armenia.\\
$^{3}$Yerevan Physics Institute,
Alikhanian Brothers St. 2, Yerevan 375036, Armenia.\\
$^{4}$ FOM-Instituut voor Atoom-en Molecuulfysica,Kruislaan
407,1098SJ Amsterdam, The Netherlands.\\
$^{*}$ Corresponding author: gevorkia@phys.sinica.edu.tw }

\begin{abstract}

{\bf Abstract}

\vskip 0.3 truecm

We show that the numerical code used in the above mentioned paper
does not take into account the multiple scattering effects of
electromagnetic field properly and is therefore incorrect.

\end{abstract}


\maketitle

Radiation originating from relativistic charged particle passing
through a random stack of foils is considered by Varfolomeev and
et al \cite{varf} using some numerical code. We treated the same
problem  analytically \cite{ZHG,ZHVER} before. The results of this
numerical simulation qualitatively coincide with the part of our
analytical consideration representing the contribution of single
scattering to the radiation intensity\cite{ZHG}.

One of the main assumptions of the code used in \cite{varf} is the
constancy of the transversal to foils component of the photon
momentum. Writing this condition explicitly for two consecutive
foils one finds
$\sqrt{\varepsilon_i}\sin\theta_i=\sqrt{\varepsilon_{i+1}}\sin\theta_{i+1}$,
where $\varepsilon_i$ and $\theta_i$ are the dielectric constant
and the angle of the photon momentum with the normal to foils in
the $i-th$ foil. One immediately recognizes in this condition the
Fresnel law of refraction. As it is well known  Fresnel laws of
reflection and refraction are correct in the geometrical optics
limit when photons are treated as rays. Therefore the claim of
paper \cite{varf} that their code is universal and applicable for
all photon wavelengths is incorrect.

The condition for applicability of geometrical optics
approximation is that the photon wavelength is smaller than all
characteristic sizes of the system. In our case the characteristic
size is the foil thickness $d$. The corrections to geometrical
optics  when considering scattering on a single foil are of order
$ \lambda/d$. However for a multiple foil system these corrections
become of order $N\lambda/d$ and can be large even for
$\lambda<<d$ if number of foils $N>>1$. Therefore using
geometrical optics approximation for solving Maxwell equations in
the multiple stack even for $\lambda<<d$, as in \cite{varf}, is
incorrect.

As we mentioned above the numerical results of \cite{varf}
qualitatively reproduce the results of single scattering
contribution to radiation intensity of our analytical
consideration \cite{ZHG}. This means that the contribution of
multiple scattering effects of electromagnetic field to the
spectral-angular radiation intensity within the geometrical optics
approximation is negligible compared to the single scattering
contribution which is simply the incoherent sum of transition
radiation photons from random interfaces. However if one properly
takes into account the multiple scattering effects their
contribution to radiation intensity can exceed the single
scattering contribution as it is shown in \cite{ZHG}.

Note that we also in our analytical consideration \cite{ZHG,
ZHVER} use the geometrical optics approximation. However we only
use this to consider of single scattering by a foil in order to
find the transmission coefficient through a foil. But we do not
use it  for solving whole Maxwell equations for a random stack as
in \cite{varf}. Note also that in our analytical consideration
\cite{ZHG} the transversal component of photon momentum in
contrary to \cite{varf} is not a constant but is a variable
integrated quantity.

 It is obvious that the angular distribution of emitted photons in
the weak absorbing multiply-scattering random layered stack should
be the same in the forward and backward directions far away from
the radiating system no matter what is the origin of appearing
photons: Cherenkov radiation, transition radiation or other
mechanisms. Even if the photons initially appear on the forward to
particle velocity direction as, for example in the Cherenkov
radiation case, after several scattering events on the random
spaced interfaces the number of backward and forward photons will
become approximately equal. Note that if this was not so we could
not observe such a fundamental phenomena as diffuse scattering of
light. Instead the numerical simulation shows exceeding of forward
intensity up to four order even in the absence of absorption.  Of
course it is very difficult to achieve the photon multiple
scattering regime $\lambda<<l<<l_{in},L$, where $l,l_{in}$ are
elastic and inelastic mean free paths of photon in the medium,
respectively and $L$ is the  size of the system, in the X-ray
region. However we think that in some special cases it is still
possible for UV and soft X-rays \cite{ZHVER} . At the same time it
is much more easy to achieve photon multiple scattering regime in
the optical region that we have demonstrated experimentally
\cite{PRL}. Summarizing, the code used in \cite{varf} to consider
the radiation of charged particles in random stack reproduces
qualitatively only single scattering contribution of our
analytical consideration and does not take into account  multiple
scattering effects properly. Therefore in all those cases where
the photon multiple scattering effects are important it is
incorrect.

The problem of numerical simulation of radiation of a relativistic
particle passing through a random stack remains a difficult and
open problem. The paper of Varfolomeev and et al \cite{varf}
contributes to a better understanding of problem.  We hope that
authors will take into account these comments  and will come up
with the revised simulation.


\begin{thebibliography}{99}

\bibitem{varf} A.A.Varfolomeev, M.J.van der Wiel, T.V.Yarovoy and D.A.Ovchinnikov,NIM B
{\bf 256},705,(2007).
\bibitem{ZHG} Zh.S.Gevorkian, Phys.Rev.E {\bf 57}, 2338 (1998); Sov.Phys.JETP {\bf 114},91, (1998).
\bibitem{ZHVER} Zh.S.Gevorkian and J.Verhoeven, NIM B {\bf 252},57,(2006).
\bibitem{PRL}Zh.S.Gevorkian and et al,Phys.Rev.Lett.,{\bf97},044801,(2006).

\end{thebibliography}
\end{document}